\documentclass[12pt]{iopart}

\usepackage[final]{epsfig}

\begin{document}

\title[Elastic Energy Loss of High-Energy Partons]{Monte Carlo Simulation for Elastic Energy Loss of High-Energy Partons in Quark-Gluon Plasma}

\author{\underline{J.~Auvinen}, K.~J.~Eskola, H.~Holopainen and T.~Renk}

\address{Department of Physics, P.O. Box 35, FI-40014 University of Jyv\"askyl\"a, Finland}
\address{Helsinki Institute of Physics, P.O. Box 64, FI-00014 University of Helsinki, Finland}
\ead{jussi.a.m.auvinen@jyu.fi}

\begin{abstract}
We examine the significance of $2 \rightarrow 2$ partonic collisions as the suppression mechanism of high-energy partons in the strongly interacting medium formed in ultrarelativistic heavy ion collisions. For this purpose, we have developed a Monte Carlo simulation describing the interactions of perturbatively produced, non-eikonally propagating high-energy partons with the quarks and gluons from the expanding QCD medium. The partonic collision rates are computed in leading-order perturbative QCD (pQCD), while three different hydrodynamical scenarios are used to model the medium. We compare our results with the suppression observed in $\sqrt{s_{NN}}=200$ GeV Au+Au collisions at the BNL-RHIC. We find the incoherent nature of elastic energy loss incompatible with the measured data and the effect of the initial state fluctuations small.
\end{abstract}

\noindent {\it 1. Introduction:} 
A substantial suppression of high-energy hadrons has been measured in $\sqrt{s_{NN}}=200$ GeV Au+Au collisions at the BNL-RHIC \cite{Adler:2006hu}. This is believed to be a consequence of the energy loss of hard partons traversing a strongly interacting medium.

To study the relevant physics of this phenomenon in as detailed manner as possible, we have developed a Monte Carlo (MC) simulation for the hard parton's interaction with the medium \cite{Auvinen:2009qm,Auvinen:2010yt,Renk:2011qi}. This is similar to the (perhaps even more ambitious) MC models JEWEL (Jet Evolution With Energy Loss) \cite{JEWEL} and MARTINI (Modular Algorithm for Relativistic Treatment of heavy IoN Interactions) \cite{MARTINI}. However, while JEWEL and MARTINI include both elastic and radiative energy-loss components, we concentrate purely on the elastic energy loss for the time being.\\

\noindent {\it 2. The model:} 
We model the elastic energy loss of a hard parton by incoherent partonic $2\rightarrow$ 2 processes in pQCD, with scattering partners sampled from the medium. Three different hydrodynamical scenarios are used to model the QCD medium: {\it i)} a (1+1)-dimensional hydro \cite{Eskola:2005ue} with initial conditions from the EKRT model \cite{EKRT} for central heavy ion collisions, {\it ii)} a (2+1)-dimensional hydro \cite{Holopainen:2010gz} with a smooth sWN profile \cite{Kolb:2001qz} obtained from the optical Glauber model for non-central collisions, and {\it iii)} an event-by-event hydro \cite{Holopainen:2010gz} with an eBC profile \cite{Kolb:2001qz} from the Monte Carlo Glauber model to study the effects of the initial state density fluctuations.

Our approach is based on the scattering rate $\Gamma_i (p_1,u(x),T(x))$ for a high-energy parton of a type $i$ with 4-momentum $p_1$, accounting for all possible partonic processes $ij\rightarrow kl$. The flow 4-velocity $u(x)$ and the temperature $T(x)$ of the medium are given by the hydrodynamical model. In the local rest-frame of the fluid, we can express the scattering rate for a process $ij\rightarrow kl$ as follows \cite{Auvinen:2009qm}:

\begin{equation}
\label{scattrate}
\Gamma_{ij\rightarrow kl} = \frac{1}{16\pi^2E_1^2}\int_{\frac{m^2}{2E_1}}^{\infty}dE_2f_j(E_2,T) \int_{2m^2}^{4E_1E_2}ds [s\sigma_{ij\rightarrow kl}(s,m^2)].
\end{equation}
Here $E_1$ is the energy of the high-energy parton $i$ in this frame and $E_2$ is the energy of the thermal particle $j$ with a distribution function $f_j(E_2,T)$, which is the Bose-Einstein distribution for gluons and the Fermi-Dirac distribution for quarks. The scattering cross section $\sigma_{ij\rightarrow kl}(s,m^2)$ depends on the standard Mandelstam variable $s$. A thermal-mass-like overall cut-off scale $m=s_m g_sT$ is introduced in order to regularize the singularities appearing in the cross section when the momentum exchange between partons approaches zero. Here $g_s$ is the strong coupling constant, which we keep fixed with momentum scale. The free parameters of our model are thus $s_m$ and $\alpha_s = \frac{g_s^2}{4 \pi}$. 

The hard parton is propagated through the plasma in small time steps $\Delta t$. At each step, the probability for a collision is given by the Poisson distribution $1-e^{-\Gamma_i \Delta t}$. We always assume there is no significant interaction between the high-energy parton and the fully hadronic medium, and thus no collisions happen in regions with temperature below the decoupling temperature $T_{\rm dec}$. The medium-modified distribution of high-energy partons obtained in the end can be convoluted with a fragmentation function to calculate the nuclear modification factor $R_{AA}(P_T,y,\phi) = \frac{dN_{AA}/dP_Tdy d\phi}{\langle N_{\rm BC}/\sigma_{NN} \rangle\,d\sigma^{pp}/dP_Tdy d\phi}.$ \\

\noindent {\it 3. Results:} 
In the following, our interest is in the high-$P_T$ neutral pions produced in $\sqrt{s_{NN}}=200$ GeV Au+Au collisions. To achieve roughly the right amount of nuclear modification in the 0-10\% centrality bin and to emulate also the incoherent higher-order processes, we set $\alpha_s = 0.5$ and $s_m=1.0$.

\begin{figure*}[htb]
\centering
\includegraphics[width=16cm]{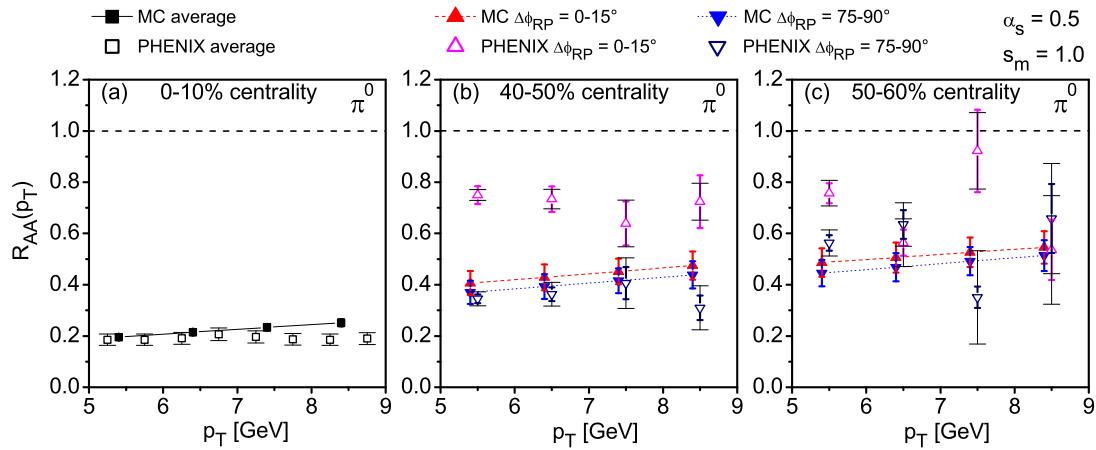}
\caption{(Color online) Left panel: The \(\pi^0\) nuclear modification factor for 0-10\% centrality, averaged over the reaction plane angle. Middle and right panel: The \(\pi^0\) nuclear modification factor dependence on the reaction plane angle $\Delta \phi$ for 40-50\% (middle panel) and 50-60\% centrality (right panel). The simulation points (solid squares and triangles) are connected with lines to guide the eye. The PHENIX data are from \cite{PHENIX-R_AA} (0-10\% centrality, open squares) and \cite{PHENIX-R_AA-RP} (40-50\% and 50-60\% centrality, open triangles). Colored bars with small cap represent statistical errors; black bars with wide cap are systematic errors.}
\label{graph_raaphi3}
\end{figure*}

The simulation results for the 0-10\%, 40-50\% and 50-60\% centrality bins, compared with the measurements, are shown in Fig. \ref{graph_raaphi3}. While the $P_T$-behaviour of the obtained $R_{AA}$ is compatible with the data within the studied transverse momentum range, it is clear from the figure that our model cannot reproduce the reaction plane angle dependence seen in the PHENIX experiment. Also the inclusive, angle-averaged nuclear modification factor fails to match with the experimental data: The computed suppression decreases too slowly as one advances to the more peripheral collisions. 

\begin{figure}[htb]
\includegraphics[width=16cm]{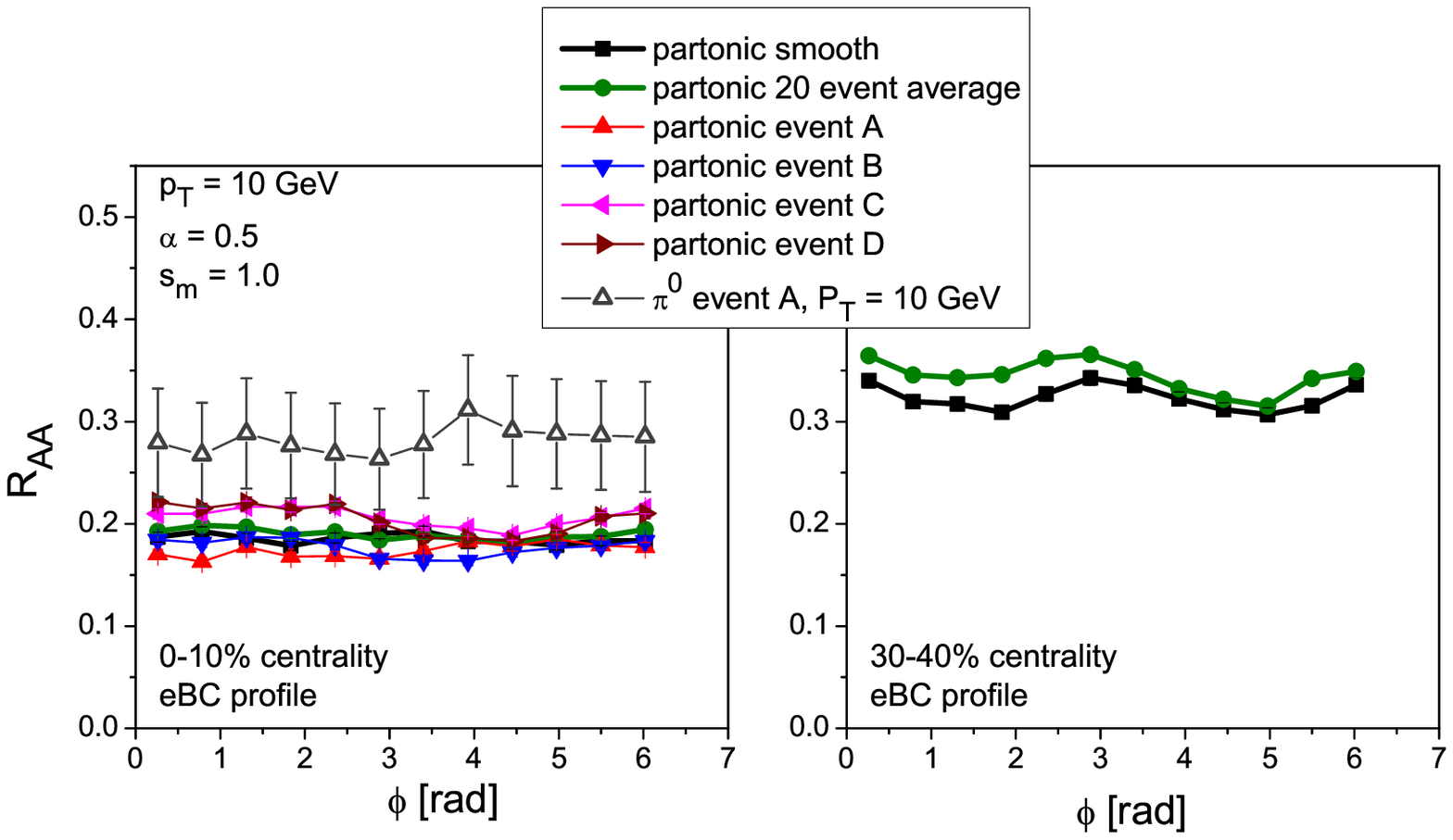}
\caption{\label{F-RAA-elastic-20evt} The partonic nuclear suppression factor $R_{AA}$ for central 200 AGeV Au-Au collisions at $p_T=10$ GeV as a function of the angle of outgoing partons with respect to the event plane. Left panel includes plots for smooth initial conditions, for four different events with fluctuating initial conditions and for an average over 20 fluctuation events in 0-10\% centrality. Nuclear suppression factor for $\pi^0$ at $P_T=10$ GeV in one event is also displayed. Right panel includes plots for smooth initial conditions and for an average over 20 fluctuation events in 30-40\% centrality.}
\end{figure}

In Figure \ref{F-RAA-elastic-20evt} we compare the angular dependence of the partonic $R_{AA}$ for fluctuating initial state geometry with the result for smooth initial conditions. While the variation in $R_{AA}$ between events can be considered notable, the angular variation within a single event is rather weak. The average over 20 events with fluctuating initial conditions, keeping the same value $\alpha_s=0.5$ for the  fluctuating and smooth cases, equals the smooth initial condition scenario with fairly good accuracy. In non-central collisions case the average of the events with fluctuating initial conditions is found to be systematically above the smooth initial conditions curve, however.\\

\noindent {\it 4. Summary:} 
Our result for $R_{AA}(\phi)$ in non-central collisions demonstrates that a purely incoherent energy-loss framework contradicts the present RHIC data. The weak sensitivity of the elastic energy loss model to the angle-dependent observables is clearly seen also in the fluctuating initial state study. In the central collisions, no difference is seen between the fluctuating and the smooth initial conditions when an average over 20 events has been taken. In the non-central collisions the fluctuating conditions do appear to produce somewhat smaller suppression compared to the smooth background.\\

\noindent {\it Acknowledgments:} 
J.A. gratefully acknowledges the grant from the Jenny and Antti Wihuri Foundation. This work was also supported by the national Graduate School of Particle and Nuclear Physics, the Academy research program of the Academy of Finland (Project No. 130472) and Academy Project 133005. CSC -- IT Center for Science Ltd. is acknowledged for the allocation of computational resources.

\pagebreak

\end{document}